\documentclass[12pt]{iopart}
\usepackage{mathdesign}
\usepackage{url}
\usepackage{pgfplots}
\usepackage{xcolor}
\usepackage{subcaption}
\usepackage[binary-units=true]{siunitx}
\usetikzlibrary{patterns}
\pgfplotsset{compat = 1.5}

\definecolor{color0}{HTML}{003F5C}
\definecolor{color1}{HTML}{7A5195}
\definecolor{color2}{HTML}{EF5675}
\definecolor{color3}{HTML}{FFA600}

\begin{document}

\title[\small{Combining processing throughput, low latency and timing accuracy in experiment control}]{Combining processing throughput, low latency and timing accuracy in experiment control}

\author{Chun Kit Lam$^1$, Stephan Maka$^1$, David Nadlinger$^{1,2}$ Chris Ballance$^2$ and S\'ebastien Bourdeauducq$^1$}


\address{$^1$ M-Labs Limited, G/F Kam Hoi Mansion, 31 Pan Hoi Street, Quarry Bay, Hong Kong.}
\address{$^2 $ Department of Physics, University of Oxford, Clarendon Laboratory, Parks Road, Oxford OX1 3PU, U.K.}
\ead{sb@m-labs.hk}

\begin{abstract}
We ported the firmware of the ARTIQ\textsuperscript{\tiny\textregistered} experiment control infrastructure to an embedded system based on a commercial Xilinx\textsuperscript{\tiny\textregistered} Zynq\textsuperscript{\tiny\textregistered}-7000 system-on-chip.
It contains high-performance hardwired CPU cores integrated with FPGA fabric.
As with previous ARTIQ systems, the FPGA fabric is responsible for timing all I/O signals to and from peripherals, thereby retaining the exquisite precision required by most quantum physics experiments.
A significant amount of latency is incurred by the hardwired interface between the CPU core and FPGA fabric of the Zynq-7000 chip; creative use of the CPU's cache-coherent accelerator ports and the CPU's event flag allowed us to reduce this latency and achieve better I/O performance than previous ARTIQ systems.
The performance of the hardwired CPU core, in particular when floating-point computation is involved, greatly exceeds that of previous ARTIQ systems based on a softcore CPU.
This makes it interesting to execute intensive computations on the embedded system, with a low-latency path to the experiment.
We extended the ARTIQ compiler so that many mathematical functions and matrix operations can be programmed by the user, using the familiar NumPy syntax.
\end{abstract}

\section{Introduction}
The ARTIQ experiment control system\cite{artiq}, initiated by the NIST Ion Storage Group, is a popular choice for many other researchers working with trapped ions\cite{lucas1}\cite{entangler}\cite{monroe}\cite{lucas2}\cite{allcock}\cite{slichter1}\cite{slichter2}, but also ultracold atoms\cite{saffman}\cite{santos}, diamond NV centers\cite{bassett}, and hybrid systems\cite{brown}\cite{kotochigova}.
One of its key features is the ARTIQ-Python language, a subset of the popular Python programming language which has been extended to provide real-time I/O features.
The subset has been chosen so the language can be compiled, whereas Python is traditionally an interpreted language.
Compilation brings significant execution performance advantages and makes execution time more deterministic.
To reduce I/O latency and make it more deterministic, the compiled ARTIQ-Python code is executed on an embedded system called the ``core device'' which is directly connected to the various peripherals controlling the experiment and collecting data, for instance, digital I/O channels, direct digital synthesizers (DDS), digital-to-analog converters (DAC), and analog-to-digital converters (ADC).
Yet, the timing accuracy of embedded code execution on a CPU with directly programmed I/O would not be sufficient for many applications that require nanosecond time resolution and sub-nanosecond (or even sub-picosecond) jitter.
Therefore, in ARTIQ, all signals are eventually timed by a custom gateware design called the ``RTIO (Real-Time Input and Output) core'', which is essentially a multichannel, modular and multi-protocol digital-to-time (DTC) and time-to-digital (TDC) core. The RTIO core is abstracted in the ARTIQ-Python language for user convenience.
For example, executing the instructions \verb!ttl.pulse(0.1*ms); delay(0.2*ms); ttl.pulse(0.3*ms)! produces the result that one would expect, with the output signal timed by gateware to sub-nanosecond accuracy.

The ARTIQ-Python language interfaces naturally with Python.
The top-level of an ARTIQ experiment is written in regular interpreted Python, which then calls an ARTIQ-Python function for real-time execution on the core device.
The ARTIQ machinery automatically compiles the ARTIQ-Python function and all other ARTIQ-Python functions that it calls (the sum of which is called a ``kernel''), uploads it to the core device, and begins its execution.
The core device may return results or execute arbitrary interpreted Python functions on the host computer thanks to a remote procedure call (RPC) mechanism, whereby calling a regular Python function from a kernel causes the core device to transfer the arguments of the call to the host, request the host to execute the function in its regular Python interpreter, and retrieve the result once execution is complete.
Communication between the host and the core device occurs over TCP/IP and Ethernet, which does not require custom host drivers and supports long cabling lengths without ground loops.

The first ARTIQ systems have been using a softcore OpenRISC (mor1kx\cite{mor1kx}) CPU to execute the compiled ARTIQ-Python code, integrated into the same FPGA as the RTIO core and the rest of the gateware.
Essential system-on-chip components such as a memory controller and bus interconnect are provided by the Migen and MiSoC packages\cite{migenmisoc}; Migen also serves as a powerful design language for the RTIO core and most other gateware components of ARTIQ.
This solution was preferred due to its simplicity, portability, consistency, and high level of control over the logic design; in particular with respect to the CPU-RTIO core interface and the possibilities to interrupt and isolate the CPU core running the user-defined kernel.
As will be evident in the rest of this paper, those points represented significant risks on Zynq-7000.

The main disadvantage of softcore CPUs is their slow execution speed, which is due to logic and routing delays inherent to FPGA technology causing a low maximum clock frequency, as well as FPGA routing limitations that preclude the efficient implementation of certain acceleration structures that are called for by some superscalar CPU designs.
In the context of ARTIQ, this limits the amount of programmable algorithmic processing that can be performed in real time in an experiment.
This work presents one solution to run kernels on a hardwired CPU core (Arm\textsuperscript{\tiny\textregistered} Cortex\textsuperscript{\tiny\textregistered}-A9), by leveraging commercial Zynq-7000 microchips from Xilinx that include, on the same device, both CPU cores and traditional system-on-chip peripherals (collectively referred to as ``PS'', which stands for ``Programmable System'') and FPGA fabric (referred to as ``PL'' for ``Programmable Logic'').

\section{Software stack and workflow}
The ARTIQ firmware is written in Rust, a modern multi-paradigm programming language designed for performance and safety.
It also uses the smoltcp TCP/IP stack, also written in Rust, that was developed by M-Labs to replace lwip within the ARTIQ project.

While porting the firmware to Zynq, we also adapted it to the latest evolutions of the Rust programming language; in particular, its implementation of cooperative multitasking is now based on the ``async-await'' asynchronous functions that have been recently introduced in Rust.
We also developed Rust drivers for the required PS peripherals, as well as a pure Rust bootloader to replace Xilinx's FSBL.
Our bootloader easily supports loading the firmware payload as well as the PL gateware from the Ethernet network, which makes it easier to use in automated tests.

As a simple demonstration of the performance of our software stack, we measured the speed of the smoltcp TCP/IP stack with async-await and our Rust Ethernet driver, by transferring raw TCP data without the ARTIQ RPC protocol.
Each buffer transferred at the Rust async-await layer is \SI{100}{\kilo\byte}.
On the other side of the TCP link, we used a Linux PC with the \verb!netcat! utility.
We measured the host-to-device throughput to be \SI{108}{\mebi\byte/\s}, and the device-to-host throughput to be \SI{102}{\mebi\byte/\s}.
These numbers are close to the hardware limit of the physical layer of Gigabit Ethernet.

The ARTIQ runtime and the PL gateware are continuously built and delivered using the Hydra\cite{hydra} continuous integration system.
The Hydra system, with some modifications of our own to escape restrictions of the Nix sandbox\cite{nix}, also runs an hardware-in-the-loop test suite after each build to help detect regressions early.
Up-to-date firmware images that are ready to be placed on a bootable SD card can be downloaded from Hydra's web-based interface.
Use of the corresponding Nix package manager\cite{nix} makes development and continuous integration builds reproducible by, among other things, pinning versions of the tools involved in the build process.
This is particularly important for Rust, which suffers from frequent incompatibility issues when using the ``nightly'' version.

\section{Dual-core architecure}
The ARTIQ softcore gateware design uses two CPU cores in order to separate kernel execution from housekeeping and other tasks.
Kernels are run on a dedicated CPU core (called the ``kernel CPU'') which is under control of the other CPU core (called ``comms CPU'').
The comms CPU is responsible for running the TCP/IP stack, starting and stopping the kernel CPU, loading kernels, transferring RPC data between the kernels and the network, and performing various auxiliary functions such as operating a built-in RTIO logic analyzer and forwarding logs to the network.
The benefits of this architecture include the ability to break infinite loops in kernels without manually resetting the core device, shielding kernel real-time performance from unpredictable network events, performing certain tasks in parallel (such as retrieving data from the built-in RTIO logic analyzer) without affecting kernel execution time, and backgrounding the sending of data to the network while kernels continue their execution (a feature called ``asynchronous RPC'').

While some Zynq chips are specified as multicore devices, they are mostly used in traditional symmetric multiprocessing (SMP) configurations with the Linux operating system, and porting ARTIQ's asymmetric dual-core architecture was a significant risk of the project.
After exploring various approaches, we used separate memory allocators per core with message passing to transfer data ownership. For kernel CPU resets, we utilize Zynq's Generic Interrupt Controller (GIC) for the comms CPU to restart the kernel CPU, with the comms CPU sending an IRQ to the kernel CPU which then performs a software reset.
While this approach would be slower due to additional copying between two cores, this greatly simplified the handling of data lifetime, as data is owned by either one of the CPU, no dangling pointer will be present after kernel CPU reset.
Unsafe code is only needed for synchronization primitives, including channels and mutexes.
Other approaches we have explored, such stack unwind from interrupt and shared allocator, are significantly more difficult and harder to maintain.

\begin{figure}[t!]
\centering
\begin{subfigure}[t]{0.5\textwidth}
\centering
\begin{tikzpicture}
\begin{axis}[
    ybar,
    bar width=.5cm,
    enlarge x limits=0.4,
    symbolic x coords={
        Bytes,
        I32 Array,
    },
    legend style={
        at={(0.5,-0.2)},
        anchor=north,
        legend columns=2,
    },
    xtick=data,
    ylabel=Throughput (MiB/s),
    ylabel shift=-7pt,
    nodes near coords={
        \scriptsize{\pgfmathprintnumber{\pgfplotspointmeta}}
    },
    legend image code/.code={%
      \draw[#1] (0cm,-0.1cm) rectangle (0.6cm,0.1cm);
    },
]
    \addplot[ybar, fill=color0] coordinates {
        (Bytes,69.38)
        (I32 Array,51.55)
    };

    \addplot[ybar, fill=color1, postaction={pattern=dots}] coordinates {
        (Bytes,62.09)
        (I32 Array,45.34)
    };

    \addplot[ybar, fill=color2, postaction={pattern=north east lines}] coordinates {
        (Bytes,2.09)
        (I32 Array,2.07)
    };

    \addplot[ybar, fill=color3, postaction={pattern=north west lines}] coordinates {
        (Bytes,2.25)
        (I32 Array,0.39)
    };

    \legend{Zynq(H2D),Zynq(D2H),mor1kx(H2D),mor1kx(D2H)}
\end{axis}
\end{tikzpicture}
\caption{\SI{1}{\mebi\byte} per request}
\label{networkperf1}
\end{subfigure}%
~
\begin{subfigure}[t]{0.5\textwidth}
\centering
\begin{tikzpicture}
\begin{axis}[
    ybar,
    bar width=.5cm,
    enlarge x limits=0.4,
    symbolic x coords={
        Bytes,
        I32 Array,
    },
    legend style={
        at={(0.5,-0.2)},
        anchor=north,
        legend columns=2,
    },
    xtick=data,
    ylabel=Throughput (MiB/s),
    ylabel shift=-7pt,
    nodes near coords={
        \scriptsize{\pgfmathprintnumber{\pgfplotspointmeta}}
    },
    legend image code/.code={%
      \draw[#1] (0cm,-0.1cm) rectangle (0.6cm,0.1cm);
    }
]
    \addplot[ybar, fill=color0] coordinates {
        (Bytes,5.89)
        (I32 Array,5.55)
    };

    \addplot[ybar, fill=color1, postaction={pattern=dots}] coordinates {
        (Bytes,5.69)
        (I32 Array,4.28)
    };

    \addplot[ybar, fill=color2, postaction={pattern=north east lines}] coordinates {
        (Bytes,0.40)
        (I32 Array,0.40)
    };

    \addplot[ybar, fill=color3, postaction={pattern=north west lines}] coordinates {
        (Bytes,0.45)
        (I32 Array,0.23)
    };

    \legend{Zynq(H2D),Zynq(D2H),mor1kx(H2D),mor1kx(D2H)}
\end{axis}
\end{tikzpicture}
\caption{\SI{1}{\kibi\byte} per request}
\label{networkperf2}
\end{subfigure}
\caption{ARTIQ RPC network transfer rates. H2D = host-to-device, D2H = device-to-host (synchronous RPC).}
\end{figure}
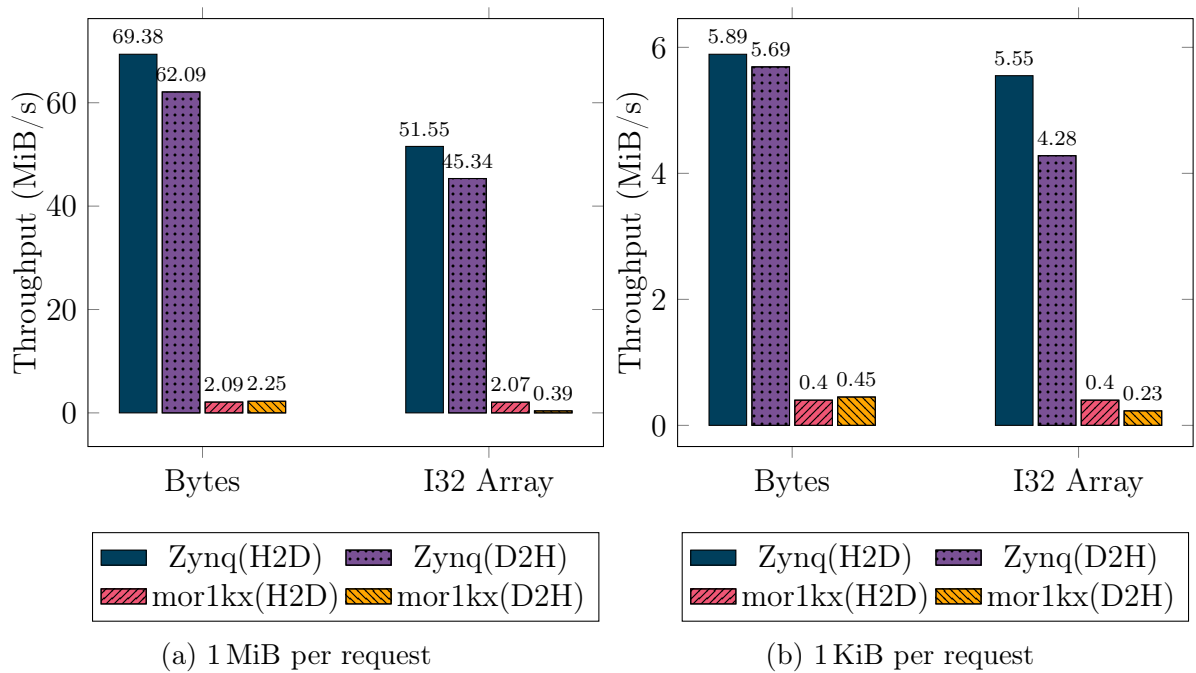

To test the performance of this design (inter-CPU communication, data encoding, network protocol, and the rest of the stack mentioned earlier), we measured the transfer speed of ARTIQ RPCs.
For this purpose, we used the \texttt{test\_performance.py} script in the ARTIQ test suite.
The hardware involved is a Zynq development kit (ZC706) with the CPU cores clocked at \SI{800}{\MHz}, and a pure FPGA Kintex-7 development kit (KC705) containing the mor1kx-based ARTIQ gateware with the CPU cores clocked at \SI{125}{\MHz}.
The same hardware is used for all the tests mentioned in this paper.
The charts in Figures \ref{networkperf1} and \ref{networkperf2} summarize the results of synchronous RPC performance of ARTIQ on Zynq comparing to previous softcore implementation with the mor1kx CPU.

The test is performed by sending or receiving a fixed size payload of different types through RPC.
For synchronous RPC, the test is repeated 10~times for \SI{1}{\mega\byte} payload, and 100~times for \SI{100}{\kibi\byte} payload, and the results are averaged to reduce the impact of jitter.
The performance of ARTIQ on Zynq is a lot better than ARTIQ with the mor1kx CPU, due to a faster CPU core and memory interface.

In general, throughput for requests with a large payload is higher than throughput for requests with smaller payload.
When the payload is small, the network latency and computation overhead of an RPC request are relatively higher.
Sending bytes is faster, as the protocol is optimized better for simpler types, and endianness conversion is not required.

For asynchronous RPC, as the kernel would not wait for response from the host, it suffers less from latency issue.
The throughput of asynchronous RPC is around \SI{69}{\mebi\byte/\s} for both \SI{1}{\mebi\byte} payload and \SI{1}{\kibi\byte} payload, making it ideal for sending experimental data back to the host.

\section{RTIO kernel initiator}
In ARTIQ, the ``RTIO kernel initiator'' is a gateware component that interfaces the kernel CPU to the RTIO core and allows submission of individual events.

\begin{figure}
\centering
\begin{tikzpicture}
\begin{axis}[
    ybar,
    width=.8\textwidth,
    bar width=1cm,
    symbolic x coords={
        Zynq(CSR),
        Zynq(ACPKI),
        mor1kx
    },
    ymin=0,
    xtick=data,
    ylabel=Interval(ns),
    nodes near coords={
        \scriptsize{\pgfmathprintnumber{\pgfplotspointmeta}}
    },
]
    \addplot coordinates {
        (Zynq(CSR),725)
        (Zynq(ACPKI),376)
        (mor1kx,418)
    };
\end{axis}
\end{tikzpicture}
\caption{Sustained RTIO Event Interval (lower is better)}
\label{RTIO}
\end{figure}
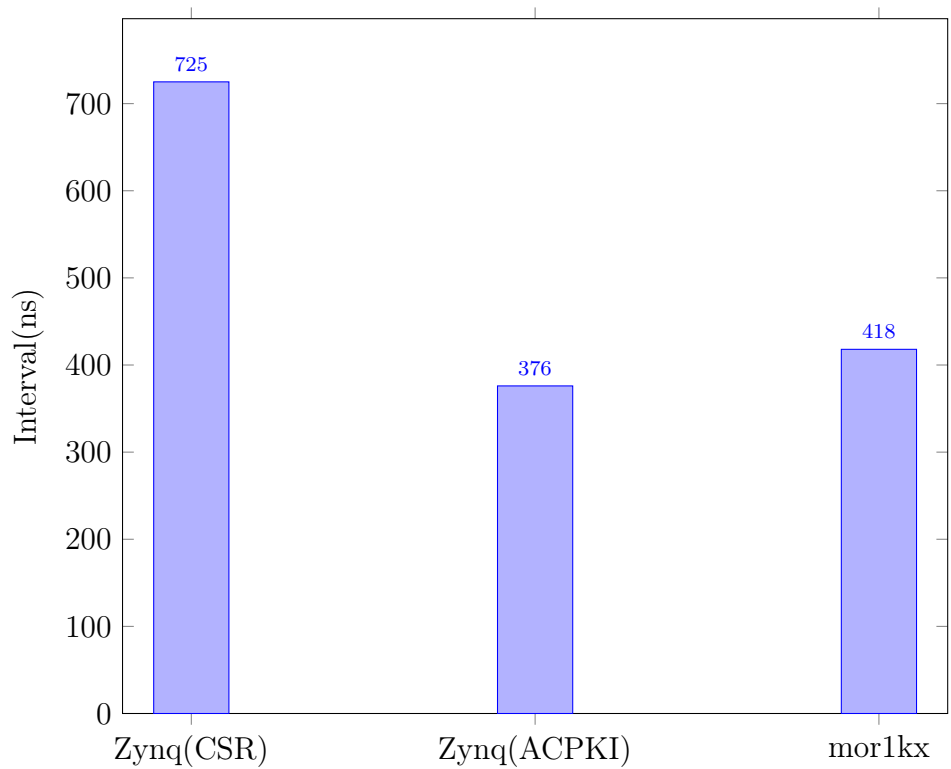

The simplest way to implement it on Zynq is use a memory-mapped configuration and status register (CSR) mechanism, which is very common in system-on-chips and also what is being used for the RTIO kernel initiator in the softcore ARTIQ design.
This can be done by using the general-purpose (GP) AXI master ports of the PS/PL interface, and is automated and integrated with Migen and MiSoC by the migen-axi library\cite{migenaxi}.
Unfortunately, this approach yields poor performance due to very high-latency clock domain transfer circuits for the AXI general purpose ports.
One typical AXI GP transaction takes 72 CPU cycles, and programming one RTIO event requires 7 such transactions.
Those circuits are part of the on-chip PS/PL interface hardware, and cannot be modified.
As a result, Zynq becomes over 70\% slower than mor1kx at submitting simple RTIO events (Figure \ref{RTIO}).

We discovered through experimentation that the cache coherent accelerator (AXI ACP) ports have lower latency than the AXI GP ports.
However, these ACP ports needs to be mastered by the PL; in other words, AXI ACP transfers can only be initiated by the PL and not the CPU.
These ACP ports simply give the PL a view into the PS memory space.
We needed an additional low-latency mechanism to cause the PL to read the PS memory for a RTIO event to be submitted.
Further experimentation revealed that the \verb!evento! lines of the PS, toggled by the Arm \verb!sev! instruction, are usable for this purpose.
Their latency is very low, and we speculate that the logic level generated inside the CPU core is simply transferred into the PL clock domain with a 2-stage flip-flop synchronizer.
We modified our firmware to avoid all uses of the \verb!sev! instruction, and repurposed the instruction to signal the submission of a new RTIO event.
We designed a new kernel initiator core, called ``ACPKI'', that monitors the \verb!evento! line, and reads the PL memory every time the \verb!evento! level changes to retrieve the RTIO event to be submitted, then writes back the transaction status.
The writing back of the status is done by the ACPKI core over the same AXI ACP port.
The AXI ACP ports implement cache coherency, so the CPU can write an event and poll memory for the transaction status without incurring a huge penalty from cache invalidations.
This creative use of Zynq implementation details allows the design to exceed the RTIO event submission performance of the mor1kx-based system, as shown in Figure \ref{RTIO}.

\section{RTIO DMA and analyzer}
The RTIO DMA feature of ARTIQ allows pre-recording a fixed sequence of RTIO events, and playing it back once or several times later.
The playback is done by gateware and can achieve higher throughput than the kernel CPU.
The RTIO analyzer stores all RTIO events, originating from both the kernel CPU and the DMA gateware core.
The events from the analyzer can be later retrieved, converted into a VCD file, and viewed with tools such as GtkWave for debugging purposes.

Both DMA and the analyzer use central (DDR SDRAM) memory bandwidth.
On Zynq, we access this memory using the AXI HP ports, which are optimal for bulk, pipelined, latency-insensitive transfers.
Furthermore, due to the structure of SDRAM, sequential transfers (such as those performed by DMA and the analyzer) provide superior efficiency and throughput.

The Zynq memory system and AXI interconnect provide the expected performance here, and we are able to submit a TTL event every \SI{32}{\ns} from DMA while the analyzer is also running (and sharing memory bandwidth).
The \SI{32}{\ns} limit is coming from our DMA and RTIO implementation and not from the memory system.

\section{New compiler functions}
The NumPy~\cite{numpy} library is widely used for scientific computing using Python, including many quantum science applications.
Previously, the slow speed of floating-point computations on the softcore CPU device meant any non-trivial computations necessary for ARTIQ experiments would need to be performed via RPC on the host CPU, incurring a latency penalty for the network round-trip.
The low performance of floating-point computations was due to the fundamental limitations of softcore CPUs mentioned earlier, but also because our softcore CPU did not include a gateware floating-point unit (while it is technically feasible to implement one).
As both Cortex-A9 cores on the Zynq-7000 chip include a hardware floating point unit, are clocked faster, and implement advanced processing speed optimization techniques, it is now possible to run significant computations directly as part of the core device kernels.
This local processing approach has been pioneered by the M-ACTION control system\cite{maction}; however, in M-ACTION, I/O events are not directly handled by the PS, and M-ACTION does not have the seamless integration of processing and I/O in a single language that ARTIQ supports.

To this end, we have extended the ARTIQ-Python language with multi-dimensional arrays, using a syntax compatible with NumPy.
Element data type and number of dimensions are represented in the language type system and are required to be known during compilation, while the elements themselves are stored in dynamically allocated stack memory, with lifetime checking performed by the compiler (similar to how \texttt{list}s are also handled in ARTIQ-Python), allowing the number of elements along each axis to be specified at runtime.

Common operations are available with the same semantics as in NumPy, including element-wise unary and binary operations ($+$, $-$, $\times$, $/$, etc.), broadcasting of scalar operands, as well as matrix multiplication.
Non-trivial array views, such as produced by multi-dimensional slicing and transposition, are not currently supported (although they could easily be implemented in the future, at a minor performance cost, by extending the array representation by a variable indexing stride per axis).

The ARTIQ runtime now also provides access to a number of commonly used special functions (e.g. \texttt{sin}, \texttt{exp}, \texttt{erf}), including support for element-wise operation on arrays. The functions are made available through the same interface that users are familiar with from NumPy/SciPy on host Python (e.g.~\texttt{numpy.sin}, \texttt{scipy.special.erf}), which also facilitates sharing of numerical subroutines between code for core device and host CPUs.

The ARTIQ compiler internally lowers array operations to function calls, for which implementations are generated on-the-fly as needed.
This way, all possible combinations of element types and dimensions are supported, relying on the loop optimizer in the LLVM compiler backend for performance, while also making it easy to manually provide implementations for any particular use case.
This could be used in the future to provide particularly optimized implementations as part of the ARTIQ runtime, for instance by linking against a finely-tuned BLAS library, or a vectorized implementation of trigonometric functions.

Figure \ref{numpyperf1} shows the performance of common builtin operations on large arrays of different data types.
We measured the time required for vector addition, multiplication, and matrix multiplication on a vector of \num{4096} elements (matrix of size $64 \times 64$).
Figure \ref{numpyperf2} shows the performance of common math functions.
We measured the time required to perform those operations on an array of \num{10000} elements, and get the average time on each element. The test data is drawn uniformly from $[-2\pi, 2\pi]$ for $\sin$, $\cos$ and $\tan$, and $[0, 1]$ for the other functions.

In summary, significant numerical processing power is now available for real-time data in ARTIQ-based experiments, with latencies in the low microseconds, rather than the approx.~millisecond required for a roundtrip to the host computer.
This is beneficial for a number of common scenarios in quantum information experiments, for instance to implement adaptive sensing protocols optimized for information gain, or optimized qubit state readout schemes.

\begin{figure}[t!]
    \begin{subfigure}[t]{0.65\textwidth}
        \begin{tabular}{|l|r|r|r|}
            \hline
            \multicolumn{4}{|c|}{Total Time (ns)}\\
            \hline
            Test & Int32 & Int64 & Float \\
            \hline
            Vector Addition       & 11485   & 23414   & 40011   \\
            Vector Dot Product    & 8212    & 90636   & 31543   \\
            Matrix Multiplication & 1389867 & 4352468 & 2118920 \\
            \hline
        \end{tabular}
        \caption{Builtin-function for vector computation\\(\num{4096} elements, $64\times64$ Matrix)}
        \label{numpyperf1}
    \end{subfigure}%
    ~
    \begin{subfigure}[t]{0.35\textwidth}
        \begin{tabular}{|l|r|}
            \hline
            Function & Time (ns) \\
            \hline
            sin  & 169 \\
            cos  & 171 \\
            tan  & 279 \\
            asin & 737 \\
            acos & 729 \\
            atan & 169 \\
            exp  & 184 \\
            log  & 190 \\
            sqrt &  44 \\
            \hline
        \end{tabular}
        \caption{Common math function performance}
        \label{numpyperf2}
    \end{subfigure}
    \caption{Computation Performance}
\end{figure}

\section{Conclusion}
This work demonstrates the feasibility of using Xilinx Zynq-7000 system-on-chips as ARTIQ core devices, the performance gains that can be obtained from their CPU+FPGA architecture, and new high-level experiment control system features that become practical thanks to the higher processing power.

One limitation of these commercial devices is the high latency between the PS and PL.
This performance metric is not precisely specified by vendors; future work could focus on evaluating other commercial devices such as Xilinx Zynq Ultrascale+ and Intel\textsuperscript{\tiny\textregistered} SoC -- or, more ambitiously, developing ARTIQ core device ASICs leveraging RISC-V CPU cores.
Another area of improvement could be changing the design of the RTIO event submission mechanism to make it more insensitive to the latency.
For example, the CPU could submit several events without obtaining feedback from the gateware after each event.
This can be achieved by having a write FIFO buffer that can contain multiple events that have been submitted by the CPU.
The CPU would maintain a counter whose value is a lower bound of how many FIFO entries are available; this counter would be stored inside the PL (e.g. the CPU cache) with a low latency.
The CPU would keep writing events without querying the gateware as long as the counter is positive, decrement it for each written event, and only obtain a new counter value from the gateware after its copy has reached zero.
A similar mechanism is already in place in ARTIQ to hide the communication latency with remote RTIO cores in DRTIO systems.

The Zynq Ultrascale+ devices are particularly interesting, as they are the basis of Xilinx's RFSoC devices that also integrate up to 16 channels of 10GSPS DACs and 8 channels of 5.0GSPS ADCs (or 16 channels of 2.5GSPS ADCs) on the same chip.
The latency of the DACs and ADCs is not characterized by Xilinx, but it appears to be dozens of nanoseconds\cite{rfsoclatency}.
This makes them a suitable hardware platform for controlling superconducting qubits with ARTIQ where speed is critical; fast paths through the gateware without involving the PS could be implemented for the most latency-sensitive functions\cite{entangler} -- a form of FPGA acceleration.
The high channel densities and low power consumption that are achievable with such a solution would make it particularly attractive.

\section*{Acknowledgements}
This work was funded by the National Institute of Standards and Technology, US Department of Commerce, under contract number 1333ND19PNB680914.

We thank David Allcock for manuscript comments.

M-Labs Limited is a commercial provider of equipment and services based on ARTIQ.

\end{document}